\documentclass[pra,showpacs,showkeys,amsfonts,amsmath]{revtex4}
\usepackage{bm}
\usepackage{graphicx}
\RequirePackage{mathptm}

\numberwithin{equation}{section}
\newcommand{\si}[1]{\sigma_{#1}}
\newcommand{\sa}[2]{\sigma_{#1}^{#2}}

\newcommand{\ro}{\rho}

\newcommand{\Ga}{\Gamma}

\newcommand{\I}{\mathbb I}
\newcommand{\ket}[1]{|{#1}\rangle}

\newcommand{\C}{\mathbb C}

\newcommand{\tr}{\mathrm{tr}\,}

\newcommand{\DS}{\displaystyle}
\begin{document}
\title{Local aspects of disentanglement induced by spontaneous emission}
\author{Anna Jamr{\'o}z\footnote{
e-mail addres: ajam@ift.uni.wroc.pl}} \affiliation{Institute of
Theoretical Physics\\ University of
Wroc{\l}aw\\
Pl. M. Borna 9, 50-204 Wroc{\l}aw, Poland}

\begin{abstract}
We consider spontaneous emission of two two-level atoms interacting
with vacuum fluctuations. We study the process of disentanglement in
this system and show the possibility of changing disentanglement
time by local operations.
\end{abstract}

 \maketitle
\section{Introduction}
Creation of the entangled quantum states and its ability to transmit
information is the base for quantum information \cite{NC}. Recently, the
subject has been intensively studied basically for two reasons. Firstly, the
understanding of entanglement creation gives deep insight into the
quantum mechanics foundations, and secondly,  possible applications
in quantum cryptography, quantum computing or teleportation of
states are very promising. Hence, one of the main aims of such study
is to get  the knowledge about the complex nature of entanglement
and its evolution in time.
\par
In practice, every quantum system is open and susceptible to
interaction with its environment. This may lead to the dissipation
and destruction of correlations. Due to that, entanglement may
disappear even though the system was initially in the entangled
state. To control the process of disentanglement it is important to
preserve as much entanglement as possible, because if entanglement
once has been lost, it cannot be restored by local operations.
\par
Spontaneous emission in two-atomic systems is an example of such
noise which can diminish entanglement\cite{FT2}. On the other hand, due to the
possible photon exchange between  atoms, even in that case some
separable initial states can evolve to entangled states \cite{JJ1}. In
particular, when the interatomic distance is very small (compared
with radiation wave length), the produced entanglement remains
non-zero also for asymptotic states \cite{J}.
\par
In the present paper we study the simpler model of  two atoms
situated in  independent thermostats at zero temperature. Since the
atoms are separated by large distance, dipol - dipol interaction and
photon exchange between atoms are negligible. Reduced dynamics of
the system is given (in the Markovian approximation) by the semi -
group of completely positive linear mappings \cite{A} with generator $L$
parametrized only by the spontaneous emission rate of the single
atom. In this model, the dynamics brings all initial states into
unique asymptotic pure state, in which two atoms are  in their
ground states. Contrary to the infinite temperature case considered
in \cite{JJ}, where the neighbourhood of the asymptotic state contains only
separable states, this asymptotic state lies on the boundary of the
set of all states and there are separable as well as entangled
states which are close to it. So there are initial entangled states
which need only finite time to become separable during the evolution
(they have \textit{finite disentanglement time}), and on the other
hand some initial states disentangle asymptotically (they have
\textit{infinite disentanglement time}).
\par
The main goal of the present paper is to study local aspects of the
process of disentanglement induced by spontaneous emission. We
address the following question: what influence on the process of
disentanglement can \textit{local} operations performed on
initial states have? As we show, local operations can change the
robustness of initial entanglement against the noise, leading in
some cases to enlarging the time needed to disentangle the state. In
some classes of pure states, simple local operation can even change
this time from finite to infinite. The same is true for Werner
states. (Similar phenomenon was studied in \cite{E1}).
\par
We consider also non-local properties of quantum states which are
manifested by violation of Bell - CHSH inequalities. As we show,
during the evolution of the system initial states violating these
inequalities become local after the finite time, even if
disentanglement time is infinite.
\section{Spontaneous emission and evolution of entangled two-atomic
systems} Let us consider two two-level atoms $A$ and $B$. 
Their excited states $\ket{1}_{A(B)}$ and ground
states $\ket{0}_{A(B)}$ we
identify with vectors $e_{1}=\left(\begin{array}{c}
1\\
0\\
\end{array}\right)$ and $e_{2}=\left(\begin{array}{c}
 0\\
 1\\
\end{array}\right)$ in $\C^{2}$.
Since the atoms are separated by the long distance, it can be
assumed that they are located inside two independent environments.
The time evolution of density matrix of the considered system can be
described by the master equation \cite{A}:

\begin{eqnarray}
\frac{d\varrho}{dt}=L\varrho\,=\,\,\frac{\Ga_{AA}}{2}
\,[\,\,2\,\sigma_{-}^{A}\,\varrho\,\sigma_{+}^{A}\,
-\,\sigma_{+}^{A}\,\sigma_{-}^{A}\,\varrho\,-
\,\varrho\,\sigma_{+}^{A}\,\sigma_{-}^{A}\,]+\,\frac{\Ga_{BB}}{2}
\,[\, \,2\,\sigma_{-}^{B}\,\varrho\,\sigma_{+}^{B}\,-\,
\,\sigma_{+}^{B}\,\sigma_{-}^{B}\,\varrho\,-
\,\varrho\,\sigma_{+}^{B}\,\sigma_{-}^{B}\,] \label{generator}
\end{eqnarray}
with  definitions
$$
\sa{\pm}{A}=\si{\pm}\otimes\I,\quad
\sa{\pm}{B}=\I\otimes\si{\pm},\quad \sa{3}{A}=\si{3}\otimes
\I,\quad \sa{3}{B}=\I\otimes \si{3},\quad \si{\pm}=\frac{\DS
1}{\DS 2}\,(\si{1} \pm \,i \si{2})
$$
\\
In the following we consider two identical atoms, so
$\Ga_{AA}=\Ga_{BB}=\Ga_{0}$, where $\Ga_{0}$ is the single atom
sponatneous emission rate. Equation (\ref{generator}) can be used to
obtain the equation of matrix elements of density matrix with
respect to the basis $|1\rangle\otimes|1\rangle$, \,
$|1\rangle\otimes|0\rangle$, \, \,$|0\rangle\otimes|1\rangle$, \,
\,$|0\rangle\otimes|0\rangle$
\begin{eqnarray}
\ro_{11}(t)&&=e^{-2\Ga_{0}t}\,\ro_{11}(0)\\
\ro_{12}(t)&&=e^{\frac{3}{2}\,\Ga_{0}t}\,\ro_{12}(0)\\
\ro_{13}(t)&&=e^{\frac{3}{2}\,\Ga_{0}t}\,\ro_{13}(0)\\
\ro_{14}(t)&&=e^{-\Ga_{0}t}\,\ro_{14}(0)\\
\ro_{22}(t)&&=e^{-\Ga_{0}t}\,(\ro_{22}(0)+\ro_{11}(0))-e^{-2\Ga_{0}t}\,\ro_{11}(0)\\
\ro_{23}(t)&&=e^{-\Ga_{0}t}\,\ro_{23}(0)\\
\ro_{24}(t)&&=e^{-\frac{1}{2}\,\Ga_{0}t}
\,(\ro_{13}(0)+\ro_{24}(0))-e^{-2\Ga_{0}t}\,\ro_{13}(0)\\
\ro_{33}(t)&&=e^{-\,\Ga_{0}t}\,(\ro_{33}(0)+\ro_{11}(0))-e^{-2\Ga_{0}t}\,\ro_{11}(0)\\
\ro_{34}(t)&&=e^{-\frac{1}{2}\,\Ga_{0}t}
\,(\ro_{12}(0)+\ro_{34}(0))-e^{-\frac{3}{2}\,\Ga_{0}t}\,\ro_{12}(0)\\
\ro_{44}(t)&&=1+\,e^{-2\Ga_{0}t}\,\ro_{11}(0)-\,e^{-\Ga_{0}t}
\,(1-\ro_{44}(0)+\ro_{11}(0))
\end{eqnarray}
The remaining matrix elements can be obtained using the hermiticity
condition $\rho_{kj}=\overline{\rho}_{jk}$. In this model the
relaxation process brings all initial states to the unique
asymptotic state when both atoms are in their ground states.
\begin{equation}
\ro_{\infty}=\begin{pmatrix} 0&0&0&0\\
0&0&0&0\\
0&0&0&0\\
0&0&0&1
\end{pmatrix}\label{asstate}
\end{equation}
\par
As a measure of the amount of entanglement of the given state of
compound system $AB$, we take entanglement of formation
\cite{Bennett} which for for mixed states is given by
\begin{equation}
E\,(\,\rho\,)\,=\min\,\sum_{k}\,\lambda_{k}E\,(\,P_{k}\,),
\end{equation}
where the minimum is taken over all possible decompositions
\begin{equation}
\rho\,=\,\sum_{k}\,\lambda_{k}\,P_{k}
\end{equation}\\
In the special case of four-level systems, $E(\rho)$ is the function
of another useful quantity $C(\rho)$ called \textit{concurrence},
which we use here as a measure of entanglement \cite{HW,Woo}. The
concurrence is defined as
\begin{equation}
C=\max \left(0,\,
\sqrt{\lambda_{1}}-\sqrt{\lambda_{2}}-\sqrt{\lambda_{3}}-\sqrt{\lambda_{4}}\right)
\label{crot}
\end{equation}
where $\lambda_{i}$ are eigenvalues of the
matrix\\
\begin{equation}
\hat{\rho}=(\,\rho^{\frac{1}{2}}\,\widetilde{\rho}\,\rho^{\frac{1}{2}}\,)^{\frac{1}{2}}
\end{equation}
with $\widetilde{\rho}$ given by
\begin{equation}
\widetilde{\rho}=(\sigma_{2}\otimes\sigma_{2})\bar{\rho}(\sigma_{2}\otimes\sigma_{2})
\end{equation}
where $\bar{\rho}$ is the complex conjugation of the matrix
$\rho$. The range of concurrence is from 0 for separable states,
to 1 for
maximally entangled pure states.\\
In general concurrence is difficult to calculate analytically, so we
consider the class of density matrices $\ro$
\begin{equation}
\ro=\begin{pmatrix}
\ro_{11}&0&0&\ro_{14}\\
0&\ro_{22}&\ro_{23}&0\\
0&\ro_{23}&\ro_{33}&0\\
\ro_{41}&0&0&\ro_{44}
\end{pmatrix}\label{inistate1}
\end{equation}
One can check that the class of states given by (\ref{inistate1}) is
invariant with respect to the time evolution considered in the
paper, and
\begin{equation}
C(\ro)=\max\,\{\,0,\,C_{1},\,C_{2}\}
\end{equation}\\
where
\begin{equation}
C_{1}=2(\,|\rho_{14}|-\sqrt{\rho_{22}\rho_{33}}\,) \label{c1}
\end{equation}
\begin{equation}
C_{2}=2(\,|\rho_{23}|-\sqrt{\rho_{11}\rho_{44}}\,) \label{c2}
\end{equation}
In particular when $\rho_{14}=0$ one can see that $C_{1}$ cannot
be positive, so only $C_{2}$, if it is positive, contributes to
the concurrence. Analogously when $\rho_{23}=0$ only $C_{1}$ can
be considered.
\par\noindent
Consider now the evolution of initial states (\ref{inistate1}). If
$\rho_{14}=0$, then
\begin{equation}
\ro_{t}=\begin{pmatrix}
\ro_{11}(t)&0&0&\ro_{14}(t)\\
0&\ro_{22}(t)&0&0\\
0&0&\ro_{33}(t)&0\\
\ro_{41}(t)&0&0&\ro_{44}(t)
\end{pmatrix}\label{inistate2}
\end{equation}
and
\begin{equation}
C(\ro_{t})=\max\,\{\,0,\,C_{1}(t)\}
\end{equation}\\
where
\begin{equation}
C_{1}(t)=2e^{-\Ga_{0}t}\left[|\rho_{14}|-
\sqrt{e^{-2\Ga_{0}t}\rho_{11}^2-e^{-\Ga_{0}t}\rho_{11}(1-\rho_{44}+\rho_{11})
+(\rho_{11}+\rho_{22})(\rho_{11}+\rho_{33})}\right]
\label{con1}
\end{equation}
In the equation (\ref{con1}), $\rho_{jk}$ denote matrix elements of
the initial state.
\par
On the other hand, if $\rho_{23}=0$, then
\begin{equation}
\ro_{t}=\begin{pmatrix}
\ro_{11}(t)&0&0&0\\
0&\ro_{22}(t)&\ro_{23}(t)&0\\
0&\ro_{32}(t)&\ro_{33}(t)&0\\
0&0&0&\ro_{44}(t)
\end{pmatrix}\label{inistate3}
\end{equation}
and
\begin{equation}
C(\ro_{t})=\max\,\{\,0,\,C_{2}(t)\}
\end{equation}\\
where
\begin{equation}
C_{2}(t)=2
e^{-\Ga_{0}t}\left[|\rho_{23}|-\sqrt{1+e^{-2\Ga_{0}t}\rho_{11}^{2}
-e^{-\Ga_{0}t}\rho_{11}(1+\rho_{11}-\rho_{44})}\right]
\label{con2}
\end{equation}

\section{Disentanglement time}
In this section we study in details the time evolution of concurrence
for some classes of initial states. Depending on the initial state,
concurrence can reach value equal to zero asymptotically or at
finite time. What is much more interesting, locally equivalent
initial states with the same concurrence can disentangle at
different times. It is even possible to change the time of
disentanglement from finite to infinite. We show that this
phenomenon happens for some classes of pure states and for Werner
states.
\subsection{Some pure initial states}
Let
\begin{equation}
\phi=\frac{1}{\sqrt{2}}(\sqrt{1+\sqrt{1-C^{2}}},0,0,\sqrt{1-\sqrt{1-C^{2}}})
\end{equation}
and $P_{\phi}$ be the corresponding projection operator
\begin{equation}
P_{\phi}=\frac{1}{2}\begin{pmatrix}
1+\sqrt{1-C^{2}}&0&0&C\\
0&0&0&0\\
0&0&0&0\\
C&0&0&1-\sqrt{1-C^{2}}
\end{pmatrix}\label{inistate1c}
\end{equation}
Then
\begin{equation}
P_{\phi}(t)=\begin{pmatrix}
\tilde{c}e^{-2\Ga_{0}t}&0&0&\frac{C}{2}e^{-\Ga_{0}t}\\
0&\tilde{c}(e^{-\Ga_{0}t}-e^{-2\Ga_{0}t})&0&0\\
0&0&\tilde{c}(e^{-\Ga_{0}t}-e^{-2\Ga_{0}t})&0\\
\frac{C}{2}e^{-\Ga_{0}t}&0&0&1-2\tilde{c}e^{-\Ga_{0}t}+\tilde{c}e^{-2\Ga_{0}t}
\end{pmatrix}
\end{equation}
where
\begin{equation}
\tilde{c}=\frac{1}{2}(1+\sqrt{1-C^{2}})
\end{equation}
By (\ref{con1}) the time evolution of this initial concurrence C is
described by the following function
\begin{equation}
C(P_{\phi}(t))=e^{-2\Ga_{0}t}(1+\sqrt{1-C^{2}})-e^{-\Ga_{0}t}(1+C+\sqrt{1-C^{2}})
\end{equation}
We see that this function becomes equal to zero at time
$t_{d}(\phi)$ (\textit{time of disentanglement}), which is given by
\begin{equation}
t_{d}(\phi)=\frac{\ln(\frac{1}{2}(1+\sqrt{\frac{1+C}{1-C}}))}{\Gamma_{0}}
\end{equation}
This time is finite for $C\in[0,1)$. When $C=1$ the states are
maximally entangled and disentangle asymptotically. On the other
hand, for locally equivalent  pure  states  $\psi=(\sigma_{1}\otimes
I_{2})\,\phi$:
\begin{equation}
\psi=\frac{1}{\sqrt{2}}(0,\sqrt{1+\sqrt{1-C^{2}}},\sqrt{1-\sqrt{1-C^{2}}},0)
\end{equation}
with projection operator
\begin{equation}
P_{\psi}=\frac{1}{2}\begin{pmatrix}
0&0&0&0\\
0&1+\sqrt{1-C^{2}}&C&0\\
0&C&1-\sqrt{1-C^{2}}&0\\
0&0&0&0
\end{pmatrix}\label{inistate2c}
\end{equation}
time evolution is given by
\begin{equation}
P_{\psi}(t)=\begin{pmatrix}
0&0&0&0\\
0&\frac{1}{2}(1+\sqrt{1-C^{2}})e^{-\Ga_{0}t}&\frac{C}{2}e^{-\Ga_{0}t}&0\\
0&\frac{C}{2}e^{-\Ga_{0}t}&\frac{1}{2}(1-\sqrt{1-C^{2}})e^{-\Ga_{0}t}&0\\
0&0&0&1-e^{-\Ga_{0}t}
\end{pmatrix}
\end{equation}
and
\begin{equation}
C(P_{\psi}(t))=e^{-\Ga_{0}t}C
\end{equation}
We see that this function asymptotically goes to zero, so we can say
that states (\ref{inistate2c}) disentangle at infinite time. Thus
we show that locally equivalent pure states with the same
entanglement behave very differently during the time evolution and
simple local unitary operation performed on initial  state can
change disentanglement time from finite to infinite.
\begin{figure}\centering\includegraphics[width=80mm]{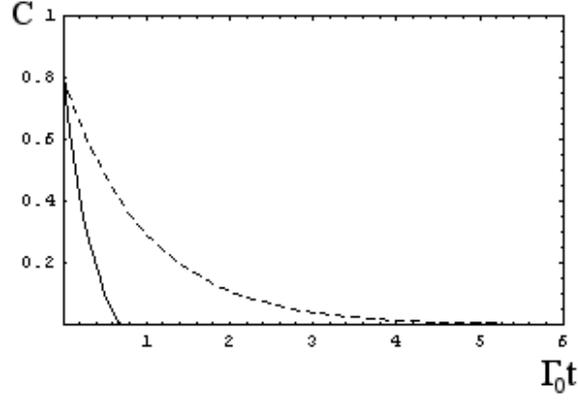}
\caption{Concurrence as function of the time for initial state (\ref{inistate2c})
(solid line) and (\ref{inistate1c}) (dotted line) with $C=\frac{4}{5}$ }\end{figure}
\subsection{Werner states}
Similar phenomenon occurs for some mixed states. Consider the class
of Werner states \cite{W}
\begin{equation}
W_{\Psi_{\pm}}\,=\,(\,1\,-\,p\,)\,
\frac{\mathbb{I}_{4}}{4}\,+\,p\,|\Psi_{\pm}\,\rangle\,\langle\,\Psi_{\pm}\,|,\label{wer1}
\end{equation}
\begin{equation}
W_{\Phi_{\pm}}\,=\,(\,1\,-\,p\,)\,\frac{\mathbb{I}_{4}}{4}\,
+\,p\,|\Phi_{\pm}\,\rangle\,\langle\,\Phi_{\pm}\,|,\label{wer2}
\end{equation}\\
\\
where $p\,\in\,[\,0,\,1\,]$ and  $\,\Phi_{\pm}\,$,
\,\,$\Psi_{\pm}$ are maximally entangled pure Bell states defined
as follows:
\begin{equation}
\Psi_{\pm}=\frac{1}{\sqrt{2}}\,[\,e_{1}\otimes e_{2} \pm e_{2}
\otimes e_{1}\,],
\end{equation}
and
\begin{equation}
\Phi_{\pm}=\frac{1}{\sqrt{2}}\,[\,e_{1}\otimes e_{1} \pm e_{2}
\otimes e_{2}\,]
\end{equation}
Since
\begin{equation}
W_{\Phi_{\pm}}(t)=\begin{pmatrix}
\frac{1}{4}e^{-2\gamma \,\,t}(1+p)&0&0&\pm\frac{1}{2}e^{-\gamma \,\,t}p\\
0&\frac{1}{2}e^{-\gamma \,\,t}-\frac{1}{4}e^{-2\gamma \,\,t}(1+p)&0&0\\
0&0&\frac{1}{2}e^{-\gamma \,\,t}-\frac{1}{4}e^{-2\gamma \,\,t}(1+p)&0\\
\pm\frac{1}{2}e^{-\gamma \,\,t}p&0&0&1-\frac{1}{2}e^{-\gamma
\,\,t}+\frac{1}{4}e^{-2\gamma\,t}(1+p)
\end{pmatrix}\label{inistate}
\end{equation}
and
\begin{equation}
C(W_{\Phi_{\pm}}(t))=e^{-\Ga_{0}t}p-\frac{1}{2}\left| e^{-2\Ga_{0}t}(1+p)-2\right|
\end{equation}
we see that
\begin{equation}
t_{d}(W_{\Phi_{\pm}})=\frac{\ln(\frac{1}{2}(\frac{1+p}{1-p}))}{\Ga_{0}}\label{time1}
\end{equation}
and this time is finite if  $p\,\in\,[\,\frac{1}{3},\,1\,]$. On the
other hand
\begin{equation}
W_{\Psi_{\pm}}(t)=\begin{pmatrix}
\frac{1}{4}e^{-2\gamma \,\,t}(1-p)&0&0&0\\
0&\frac{1}{2}e^{-\gamma \,\,t}-\frac{1}{4}e^{-2\gamma \,\,t}(1-p)&\pm\frac{1}{2}e^{-\gamma \,\,t}p&0\\
0&\pm\frac{1}{2}e^{-\gamma \,\,t}p&\frac{1}{2}e^{-\gamma \,\,t}-\frac{1}{4}e^{-2\gamma \,\,t}(1-p)&0\\
0&0&0&1-\frac{1}{2}e^{-\gamma
\,\,t}+\frac{1}{4}e^{-2\gamma\,t}(1-p)
\end{pmatrix}
\end{equation}
and
\begin{equation}
C(W_{\Psi_{\pm}}(t))=e^{-\Ga_{0}t}p-\frac{1}{2}
e^{-\Ga_{0}t}\sqrt{e^{-2\Ga_{0}t}(1-p)^{2}-4(1-p)(1-e^{-\Ga_{0}t})}
\end{equation}
When $p\,\in\,[\,\frac{1}{3},\,\frac{-1+\sqrt{5}}{2}\,)$ this
function  is monotonically decreasing  and reaches zero at time
\begin{equation}
t_{d}(W_{\Psi_{\pm}})=\frac{1}{\Ga_{0}}
\ln\frac{-1+p-\sqrt{(-1+p)^{2}p(1+p)}}{2(1-p-p^{2})}\label{time2}
\end{equation}
But when
$p\,\in\,[\,\frac{-1+\sqrt{5}}{2}\,,1],\;C(W_{\Psi_{\pm}}(t)) $ goes
to zero asymptotically, so we conclude that the disentanglement time is
infinite.
\begin{figure}
\includegraphics[width=80mm]{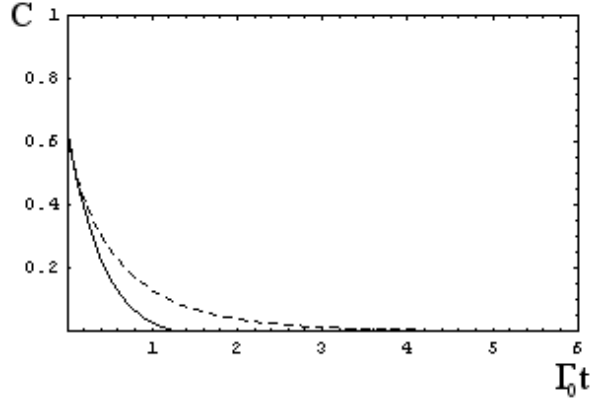}
\caption{Concurrence as function of the time for initial state $W_{\Psi_{\pm}}$
(solid line) and $W_{\Phi_{\pm}}$ (dotted line) with $p=\frac{3}{4}$}
\end{figure}
Notice that
$$
W_{\Phi_{+}}=(\I_{2}\otimes i\si{2})\,W_{\Psi_{-}}(\I_{2}\otimes
-i\si{2}),\quad W_{\Phi_{-}}=(\I_{2}\otimes
\si{1})\,W_{\Psi_{-}}\,(\I_{2}\otimes \si{1})
$$
If $p\,\in\,[\,\frac{-1+\sqrt{5}}{2}\,,1]$ then the states $
W_{\Phi_{\pm}}$ have the finite disentanglement time, whereas
locally equivalent to them $W_{\Psi_{\pm}}$ disentangle
asymptotically.
\section{CHSH inequalities}
Let us discuss now violation of Bell - CHSH inequalities by states
evolving in time. It is known that all pure states violate Bell -
CHSH inequalities whenever they are entangled. In the case of mixed
states of two two-level systems, we can apply the following
criterion \cite{HHH,HH}: Let
\begin{equation}
m(\ro)=\max_{j<k}\; (u_{j}+u_{k})
\end{equation}
where $u_{j},\, j=1,2,3$ are the eigenvalues of real symmetric
matrix $U_{\ro}$ given by
\begin{equation}
U_{\ro}=T_{\ro}^{T}\,T_{\ro}
\end{equation}
with  $T_{\ro}=(t_{nm})$,\,\,\ $t_{nm}=\tr
(\ro\,\si{n}\otimes\si{m})$. Than  $\rho$ violates  Bell - CHSH
inequalities if and only if $m(\ro)\geq1$. For the class
(\ref{inistate1})
\begin{equation}
m(\ro)=\max \left(\,u_{1}, \,u_{2}\right) \label{crot}
\end{equation}
where
\begin{equation}
u_{1}=8(|\rho_{14}|^{2}+|\rho_{23}|^{2})\label{u1}
\end{equation}
and
\begin{equation}
u_{2}=4(|\rho_{14}|^{2}+|\rho_{23}|^{2})+
\sqrt{(\rho_{23}+\rho_{32})^{2}(4|\rho_{14}|^{2}
+(\rho_{23}-\rho_{32})^{2})}+(-1+2(\ro_{11}+\ro_{44}))^{2}\label{u2}
\end{equation}
Since interaction with environment destroys correlations, we expect
that states which initially violate Bell - CHSH inequalities, during
the evolution become local i.e. non-violating these inequalities.
Consider for example pure initial states (\ref{inistate1c}). One can
check that
\begin{equation}
m(P_{\phi}(t))=\max \left(\,u_{1}, \,u_{2}\right)
\end{equation}
where
\begin{equation}
u_{1}=2e^{-2\Ga_{0}t}C^{2}
\end{equation}
and
\begin{equation}
u_{2}=e^{-2\Ga_{0}t}C^{2}+(1-2e^{-\Ga_{0}t})^{2}
\end{equation}
From the condition $m( \rho )=1$ we can calculate the
\textit{locality time } $t_{loc}$, after which Bell-CHSH
inequalities are satisfied, and we obtain

\begin{equation}
t_{loc}=\left\{\begin{array}{ll}
\frac{\ln(1+\frac{C^{2}}{4})}{\Ga_{0}}, & \qquad C\in[\,0,\,\, 2(-1+\sqrt{2})]\\
\frac{\ln2C^{2}}{2\Ga_{0}}, & \qquad C\in[\,2(-1+\sqrt{2})],\, \,1]
\end{array}\right.
\end{equation}
For all $t\in[\,0,\,t_{loc}\,]$ the states $P_{\phi}(t)$ are
entangled and violate Bell - CHSH inequalities.

On the other hand, in the case of initial states  (\ref{inistate2c})
this time  is the same. We see that even if locally equivalent
initial states disentangle at different times, the time after which
they become local is the same. Similar calculations can be done for
Werner initial states. Consider $p\,\in
\,[\,\frac{1}{\sqrt{2}},\,1\,]$. Then initial states violate Bell -
CHSH inequalities and
\begin{equation}
m(W_{\Psi_{\pm}}(t))=m(W_{\Phi_{\pm}}(t))= 2e^{-2\Ga_{0}t}p^{2}
\end{equation}
so
\begin{equation}
t_{loc}=\frac{\ln2p^{2}}{2\Ga_{0}}
\end{equation}
\section*{Acknowledgments}
The author would like to thank Lech Jakobczyk for many useful discussions and all suggestions.
This work was supported by grant IFT/W/2567.

\end{document}